\documentclass{article}

\begin{document}
\input{epsf}

\begin{center}
\Large{{\bf Universality class for bootstrap percolation with $m=3$ on the 
cubic lattice}} \\
\vskip 0.7cm
\large{N S Branco and Cristiano J Silva \\
Universidade Federal de Santa Catarina, Depto.\ de  F\'{\i}sica, \\
	88040-900, Florian\'{o}polis, SC, Brazil}

\end{center}

\begin{abstract}
    
    We study the $m=3$ bootstrap percolation model on a cubic lattice, using
Monte Carlo simulation and finite-size scaling techniques. In bootstrap
percolation, sites on a lattice are considered occupied (present)
or vacant (absent)
with probability $p$ or $1-p$, respectively. Occupied sites with less than
$m$ occupied first-neighbours are then rendered unoccupied; this culling 
process is
repeated until a stable configuration is reached. We evaluate the percolation
critical probability, $p_c$, and both scaling
powers, $y_p$ and $y_h$, and, contrarily to previous calculations, our
results indicate that the model belongs to the same universality class as
usual percolation (i.e., $m=0$). The critical spanning
probability, $R(p_c)$, is also numerically studied, for systems with 
linear sizes ranging 
from $L=32$  up to $L=480$: the value we found, $R(p_c)=0.270 \pm 0.005$, is
the same as for usual percolation with free boundary conditions.

\vskip 0.7cm


\vskip 0.7cm


\end{abstract}

\section{Introduction} \label{intro}

  	In many physical phenomena percolation effects play an important role
\cite{Perc}. Particularly, some dilute magnets are well  described, in what
concerns magnetic phase transitions, by uncorrelated diluted models
\cite{Stinch}. In these models, magnetic sites on a lattice are randomly
replaced by non-magnetic ones, and a bond links each pair of occupied
(magnetic) first neighbours. At zero temperature, the problem is purely
geometrical and  is described in the following way. Sites on a lattice are
randomly occupied with probability $p$, while all bonds are considered to be
present. A cluster is then defined as a collection of present sites which are
connected to each other through steps between occupied first-neighbours.
As $p$ increases, an infinite cluster appears for the first time at a
critical probability $p_{c}$, which is lattice dependent. In analogy with
thermal
critical phenomena, some quantities are singular at the critical point,
following a power-law behavior near $p_{c}$ \cite{Perc}. Typical examples
are the probability $P(p)$ that a site belongs to the infinite cluster,
which behaves as $P(p) \sim (p-p_c)^{\beta}$ near $p_c$, and the correlation
length $\xi(p)$, which diverges at $p_c$, according to 
$\xi(p) \sim |p-p_c|^{-\nu}$.

	Nevertheless, some systems were found to be better described by
{\it correlated} percolation models, where the presence of sites (or bonds)
depends also on their neighbourhood. Typical examples of correlated
percolation models are bootstrap percolation \cite{Corr} and
site-bond correlated percolation \cite{Rec}. The physical motivation for the
introduction of the former model comes from diluted magnetic systems where
competition between exchange (which favour a magnetic ground
state) and crystal-field (which leads to a non-magnetic ground state)
interactions takes place. To mimic this competition at zero temperature, the
bootstrap percolation model was introduced \cite{cha,NSB}: in this model,
sites on a lattice are randomly occupied with probability $p$ but only those
with at least $m$ occupied first-neighbours remain occupied. In the stable 
final configuration, all occupied sites have at least $m$ occupied
first-neighbours
or the whole lattice is empty. As our purpose in this paper is
to study one implementation of the bootstrap percolation model, we briefly
review some of its properties in what follows (for a thourough discussion
of the results up to 1990, see Reference \cite{Corr}).

	The $m=0$ case regains the usual (uncorrelated) site percolation model,
where the transition is continuous and $p_c < 1$ in two and three dimensions.
More specifically, the most precise evaluations of the critical exponents
for three-dimensional lattices
are $\nu=1/y_p=0.875 \pm 0.008$ and $\beta=0.412 \pm 0.010$ from simulation 
\cite{Ziff1} and $\nu=1/y_p=0.872 \pm 0.070$ and
$\beta=0.405 \pm 0.025$ from series \cite{series}.
In the $m=1$ case, only isolated sites are removed by the
culling process: these sites do not contribute to the critical behaviour and
$p_c$ and all critical exponents remain the same as for usual percolation. 
For $m=2$, on the other hand, isolated clusters with two sites are eliminated, 
as  well as some dangling structures of more compact clusters. The elimination 
of these dangling structures, however, does not break the infinite cluster 
(whenever present) and,
therefore, the critical probability is the same as for usual percolation.
Moreover, as the exponent $\nu$ is also connected to the formation of this 
infinite cluster, its value is the same for $m=2$ and $m=0$. In what concerns
``field'' exponents ($\beta$, for instance), previous results for
two-dimensional systems indicate a higher value of this exponent for $m=2$ 
than for usual percolation \cite{NSB2}.
Nevertheless, it has been shown later, through simulations on bigger lattices
in two dimensions \cite{belita} and general arguments applied to both two
and three dimensions \cite{Branco}, that the exponent $\beta$ is the same for
uncorrelated and $m=2$ bootstrap percolation models. 

	Let us now turn our atention
to higher values of $m$. It is generally believed that, for any value of
$m$ where only infinite clusters can survive the culling process, $p_c=1$.
This is indeed the case for $m=2d-1$ on hypercubic lattices ($d$ is the
dimension of the lattice) \cite{robert}, for $m=4$ on cubic \cite{robert} and
triangular \cite{Corr} lattices and for $m=5$ on the triangular lattice
\cite{kogut}. Moreover, it has been shown that, for these cases, the
usual finite-size scaling relation does not hold. 
This finite-size scaling predicts that, if a
suitable definition of a finite-lattice ``critical'' point, $p_{av}$, is made,
this point will approach the critical point in the thermodynamic limit,
$p_c$, as:
\begin{equation}
   p_c - p_{av} \sim L^{-1/\nu}, \label{fss1} 
\end{equation}
where $L$ is the linear size of the finite system, such that $L \gg 1$,
and $\nu$ is the usual 
critical exponent. This finite-size behaviour is indeed observed for $m \leq 2$,
but fails for high values of $m$. For $m=2d-1$ on hypercubic lattices,
it has been proven that $p_c - p_{av} \sim 1/(\log^{d-1}L)$ \cite{aizenman},
e.g., proportional to $1/\log(\log L)$ for $d=3$ dimensions. Also for
$m=4$ on the cubic lattice, the correct finite-size
behaviour is $p_c - p_{av} \sim 1/\log (\log L)$, with $p_c=1$ \cite{cerf}.
These results for high $m$ have been conjectured or tested in numerical
simulations \cite{vanenter}.

  	The results stated in the previous paragraphs do not apply to the
$m=3$ case on the cubic lattice. For this model, one expects $p_c$ to be
above the value for uncorrelated percolation (where $p_c=0.311605
\pm 0.000005$ \cite{Acha}), since the infinite cluster for usual percolation
at $p_c$ is not stable with respect to the culling process for $m=3$. On the
other hand, since finite clusters are still stable for this value of $m$
on the cubic lattice, it is expected that $p_c<1$. Numerical simulations
confirmed this scenario \cite{kogut,as}, although the relative small sizes 
used in those works indicate
that the values might not be precise (we will return to this point later).
In what concerns the critical exponents, previous results indicate that
the exponent $\nu$ is the same as for usual percolation \cite{as} but
$\beta$ is higher than its uncorrelated counterpart \cite{kogut,as}.
However, in neither of these works a extrapolation to the thermodynamic
limit is attempted, leaving the possibility that finite-size effects might
be the reason for the discrepancy in the values of $\beta$. This possibility
was first proven to be right, in the context of two-dimensional bootstrap
percolation 
models, in References \cite{belita} and later confirmed for $m=3$ on the
triangular lattice \cite{arcangelis}. The possibility of
a new universality class for bootsrap percolation with $m=3$ on the cubic
lattice is the problem we address in this work. We resort
to numerical simulation methods, which, together with finite-size scaling
analysis, allowed us to obtain more precise values for the critical parameters.
We also study the so-called critical spanning probability, i.e., the
probability that a given lattice has a cluster connecting its boundaries
at criticality \cite{langlands,Ziff2}. This quantity shows some degree of
universality: it depends on the dimension and shape of the system and on the 
specific
boundary condition but not on the lattice type (simple cubic of f.c.c., for
instance) and on the
particular kind of percolation (site or bond) \cite{Acha,Ziff2,Lin}. 
It is then interesting to
see whether it remains invariant for percolation models such that
long-range correlation is involved, like bootstrap percolation.

	The remainder of the paper is organized as follows. In the next section
we present the method and discuss some technical details, as well as the
results for the critical parameters. In Section \ref{csp} we discuss some
previous results concerning the critical spanning probability for percolation
and our results for bootstrap percolation. Finally, we summarize our results
in the last section.

\section{Method and Results} \label{mr}

      The method we use is connected to real-space renormalization-group 
and finite-size scaling procedures \cite{MCRG}. The approach needs that precise
values of the physical quantities are available; these are only obtained for
high values of the linear system size $L$. We then studied finite systems of 
size $L^3$, with $32 \leq L \leq 480$; from the results for $L \gg 1$,
it is possible to use finite-size scaling techniques to extrapolate
to the thermodynamic limit ($L=\infty$). 

\begin{figure}
\epsfxsize=9.5cm
\begin{center}
\epsffile{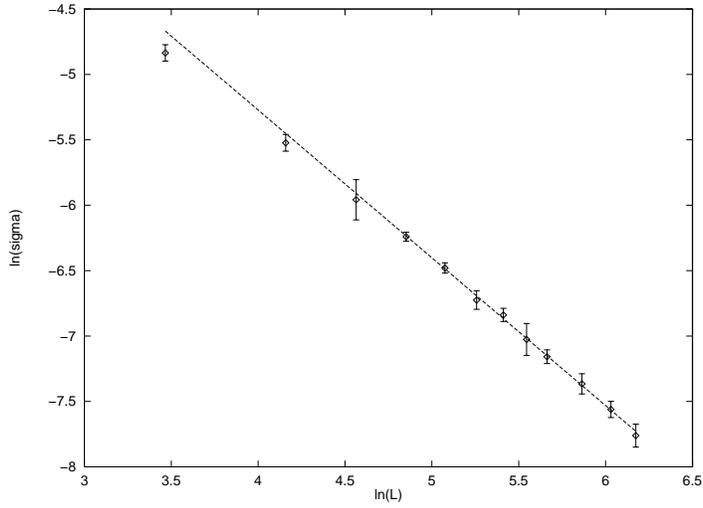}
\caption{Log-log plot $\sigma \times L$, for $32 \leq L \leq480$. The
linear size $L$ is always a multiple of 32. The dotted
line is a linear fit of the data {\it for $128 \leq L \leq 480$ only}
(see text).}
\label{sig} 
\end{center}
\end{figure}

	The critical probability $p_c$ and the critical exponent $\nu$ are
calculated as follows. For a lattice of size $L$, we occupy 
each site with probability $p$, apply the bootstrap condition and test the
lattice for percolation (here we define percolation as the presence of
a cluster which connects the bottom and top planes of the finite cubic lattice;
we discuss this and other technical points below).
Our finite-size estimate of the critical probability, $p^*$,
is taken as the value of $p$ at which the cell percolates for
the first time, when $p$ is increased from zero; 
this procedure is made for ${\cal N}$ different
runs (which correspond to ${\cal N}$ different seeds to the random number  
generator). Each run leads to a different value of $p^*$, since the lattice 
is finite. We take the average of the ${\cal N}$ values of $p^*$ as 
our estimate of $p_{av}$ (see Section \ref{intro}). It is then assumed that
$p_{av}$ will approach $p_c$ as given by Equation \ref{fss1}. Moreover, it is 
possible
to calculate the width $\sigma = \sqrt{<p^{*^2}> - p_{av}^2}$, 
which behaves as ($<p^{*^2}>$ stands for the average of  $p^{*^2}$ over the 
${\cal N}$ realizations):
\begin{equation}
   \sigma \sim L^{-1/\nu}. \label{fss2}
\end{equation}
	From the previous equation and a $\log-\log$ plot of $\sigma \times L$, 
it is then possible to obtain the value of
the critical exponent $\nu$. Tha data is depicted in Figure \ref{sig};
from the slope of the straight line we obtain
$\nu=0.89 \pm 0.04$. We compare this value with other evaluations in Table
\ref{crit}; within the numerical precision, this exponent is the same for
bootstrap percolation with $m=3$ and ordinary percolation on the cubic lattice
and agrees with previous evaluations of $\nu$ for bootstrap percolation
with $m=3$ on the cubic lattice.
Note from the graph that the straight line regime is achieved for
$L \geq 128$, while for uncorrelated percolation this regime sets in for
smaller values of $L$. This is expected, since finite-size effects are stronger 
for correlated percolation problems than for their uncorrelated counterparts
\cite{NSB,NSB3}. Therefore, we neglect the data for $L \leq 96$ and used
only lattices with $128 \leq L \leq 480$ in our linear regression analysis.

\begin{figure}
\epsfxsize=9.5cm
\begin{center}
\leavevmode
\epsffile{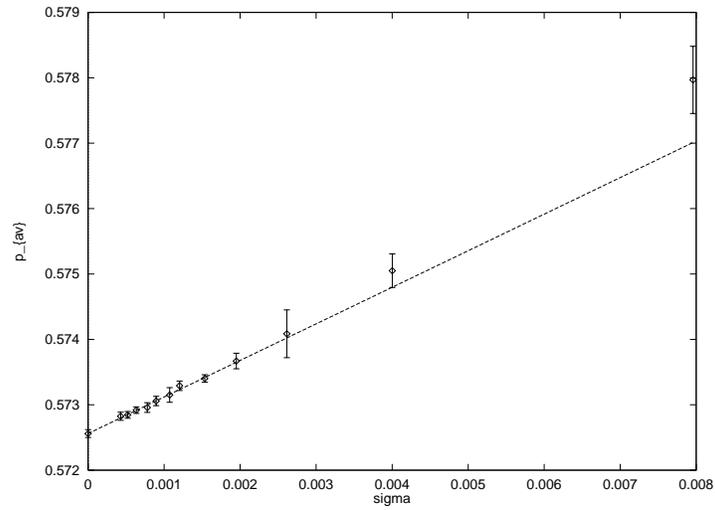}
\caption{Graph of $p_{av} \times \sigma$, for $32 \leq L \leq480$. The
linear size $L$ is always a multiple of 32. The dotted
line is a linear fit of the data {\it for $128 \leq L \leq 480$ only}
(see text).
Although not depicted in the figure, there are error bars in the
values of $\sigma$, which were taken into account in the evaluation
of $p_c$.}
\label{sigma} 
\end{center}
\end{figure}

	In order to calculate the critical threshold, we resort to Equations
\ref{fss1} and \ref{fss2}, which imply that, for $L \gg 1$:
\begin{equation}
    p_c - p_{av} \sim \sigma . \label{fss3}
\end{equation}
This is a convenient way to calculate $p_c$, since it does not depend on the
value of $\nu$; such dependence would appear if Equation \ref{fss1} was used.
It is shown in Figure \ref{sigma} a plot of $p_{av} \times \sigma$: $p_c$ is
given by the linear coeficient and the value obtained is 
$p_c=0.57256 \pm 0.00006$. This value is slightly above the one calculated in
Reference \cite{as}, in which the value of $L$ varied from 10 to 110. If we
use the same range in our calculation the extrapolated value of $p_c$ is
consistent with the result of \cite{as} (see Table \ref{crit}).

	Let us know discuss some technical points. An important
quantity is the value of $p$ at which the lattice percolates for the first
time, when $p$ is increased from zero. One needs to define what ``percolate''
means for a finite lattice. In this work, we used the rule called ${\cal R}_1$
in Reference \cite{MCRG}, i.e., a lattice percolates if, after the bootstrap 
culling process, there is a path of present sites which links the
boundaries of the lattice in a fixed direction (vertical, say). There are
other possible definitions (see \cite{MCRG}) and it is expected that all
of them lead to the
same value of $p_c$ in the thermodynamic limit. Note, however, that the
value of the critical spanning probability does depend on this definition,
as we will discuss im Section \ref{csp}. The numerical procedure we used
to test for percolation is the Hoshen-Kopelman algorithm \cite{HK}:
for usual percolation, it requires the storage of only one plane. For
bootstrap percolation, on the other hand, the bootstrap iteration needs the 
storage
of the whole lattice, due to correlation effects. To cope with this drawback,
we store the lattice in bits, instead of words: this saves memory and time,
since the updates connected to the bootstrap rule can be made in parallel for
a set of 32 sites \cite{stauffer}.
To define all six first-neighbours of sites at the boundaries of the lattice,
periodic boundary conditions are used.
It is expected that the boundary condition does not affect the
values of the critical parameters in the thermodynamic limit, since it is
a ``surface'' effect. Finally, let us
mention that the number of realizations ${\cal N}$ (see Section \ref{intro})
varied from 12000 for $L=32$ to 640 for $L=480$; the errors were calculated
as three times the standard deviation for subsets of the total number
of realizations.

\begin{table}
\caption{Critical parameters for $m=3$ bootstrap percolation and comparison
with previous results for this model and for usual percolation.}
\begin{tabular}{|l|l|l|}     \hline        
   & m=3 bootstrap percolation & usual percolation  \\ \hline\hline 
 $p_{c}$ & 0.57256$\pm$0.00006$^a$ &   \\
         & 0.5717$\pm$0.0005$^b$ & 0.311605$\pm$0.000005$^d$ \\
         & 0.568$\pm$0.002$^c$ &  \\ \hline
 $\nu$ & 0.89$\pm$0.04$^a$ & 0.875$\pm$0.008$^e$ \\
       & 0.876$\pm$0.010$^b$ & 0.872$\pm$0.070$^f$ \\ \hline
 $\beta$ & 0.37$\pm$0.03$^a$ & 0.41$\pm$0.02$^e$ \\
         & 0.6$\pm$0.1$^b$ & 0.40$\pm$0.06$^f$ \\ 
         & 0.82$\pm$0.04$^c$ & \\ \hline
\end{tabular}
\\
$^a$ Present work. $^b$ Reference \cite{as}. $^c$ Reference \cite{kogut}.\\
$^d$ Reference \cite{Acha}. $^e$ Reference \cite{Ziff1}.
$f$ Reference \cite{series}
\label{crit}  
\end{table}
      
\begin{figure}
\epsfxsize=9.5cm
\begin{center}
\leavevmode
\epsffile{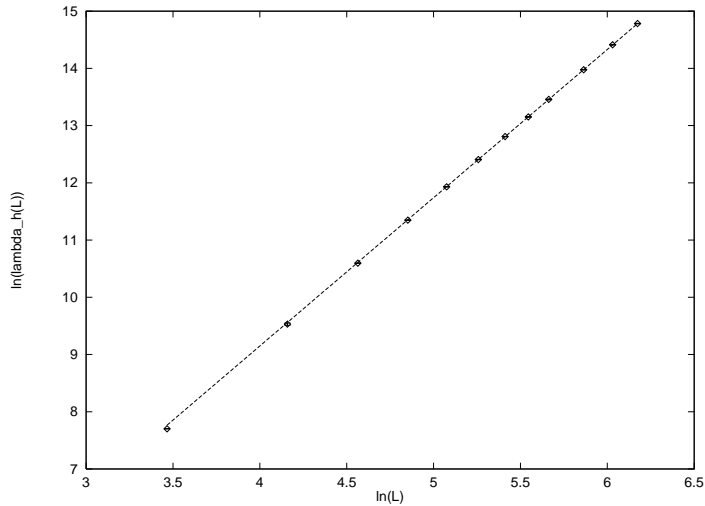}
\caption{Graph of $\log(\lambda_h(L)) \times \log(L)$, for $32 \leq L \leq480$. 
The linear size $L$ is always a multiple of 32. The dotted
line is a linear fit of the data {\it for $128 \leq L \leq 480$ only} (see text)
and $y_h$ is the angular coefficient of this line.
The error bars are smaller than the data points.}
\label{yh} 
\end{center}
\end{figure}

       To have access to the ``magnetic'' scaling power $y_{h}$, one possible
way is to define a ``ghost''- site, which is linked to all sites of the 
lattice with probability $h$ \cite{MCRG}. Within a real-space 
renormalization group framework, it is possible to calculate 
the eigenvalue $\lambda_{h}$ through
\mbox{$\lambda_{h}=<n>/p_{c}$}, where $<n>$ is the average number of occupied
sites linked to one of the boundary planes of the finite cubic lattice at the
critical threshold, $p=p_c$. The value of $p_c$ is obtained as explained above 
and is then used to calculate
$<n>$ and hence $\lambda_{h}(L)$, averaging over ${\cal M}$
configurations. The value of ${\cal M}$ varied from $\sim 320000$, for the
smaller lattices, to $\sim 20000$, for the bigger ones.
We use two procedures to obtain $y_h$ in the thermodynamic
limit. The first one is based on the fact that, for $L \rightarrow \infty$:
\begin{equation}
\lambda_h(L) = L^{y_h}, \label{fss4}
\end{equation}
with $y_h$ independent of $L$; this equation leads to
a straight line in a $\log-\log$ plot of $\lambda_h \times L$, for
$L \gg 1$. As
depicted in Figure \ref{yh}, this is indeed the case for $128 \leq L \leq 480$.
From a linear fitting of the data for this range of $L$, and resorting to
the scaling relation $\beta = (d-y_h) \nu$, we obtained the value $\beta=
0.37 \pm 0.03$ (see Table \ref{crit}). Alternatively, we could take into
account correction-to-scaling terms, through:
\begin{equation}
   \lambda_h = L^{y_h} \left( 1 + B/L \right). \label{fss5}
\end{equation}
From this equation, we see that local slopes of $\log \lambda_h \times \log L$
provide estimates of $y_h(L)$, which, when extrapolated to $L\rightarrow
\infty$, leads to an evaluation of $y_h$ in the thermodynamic limit. We
have applied this procedure for $32 \leq L \leq 480$, using three consecutive
points to calulate the local slopes and extrapolating to $L \rightarrow \infty$
through a $y_h(L) \times 1/L$ graph. The value thus obtained for $\beta$ is 
the same as for the first procedure. In
Table \ref{crit} we see that our estimate of $\beta$ disagrees with the 
values
calculated in References \cite{kogut} and \cite{as}. In evaluating this
exponent, the former uses lattices of linear size $L=35$, while the latter uses
$L=80$; neither attempted to make an extrapolation to the thermodynamic limit.
From Table \ref{crit}, we can infer that our estimate of $\beta$
for the bootstrap model we study is, within the error
bars, equal to the corresponding value of this exponent for usual
(uncorrelated) percolation. Since we have already seen that $\nu$ is also the
same for both models, we can draw the conclusion that usual percolation
and bootstrap percolation with $m=3$ on the cubic lattice belong to the
same universality class. This result contradicts References
\cite{kogut} and \cite{as}; we believe this is caused by the small lattices
used in those works. 

\section{Critical spanning probability} \label{csp}

	It has been established some time ago that the critical spanning
probability, $R(p_c)$, defined as the probability of spanning a lattice at 
the critical  point, shows some degree of universality \cite{langlands,cardy}. 
More precisely, this quantity does not depend on the lattice type and on the
kind of percolation. This result contradicts the assumption
made on early applications of real-space renormalization group
procedures to percolation. In those, it was assumed that the critical spanning
probability is equal to the critical percolation threshold, $p_c$ and,
therefore, lattice dependent \cite{MCRG}. However, later numerical tests 
confirmed and expanded the universality proposal \cite{Acha,Ziff2,Lin}.

	 Nevertheles, no study has been made on correlated models, to the best 
of our knowledge. While it is expected that {\it short-range} correlations do
not change the universality scenario \cite{NSB}, it is not clear whether 
$R(p_c)$ changes
when {\it long-range} correlations are introduced. A convenient model
to test these possibilities is the bootstrap percolation one. While the
behaviour of $R(p_c)$ is trivial for the cases where $p_c=1$, for
$m=3$ on the simple cubic lattice the presence of correlation may lead
to a non trivial behavior. We study this possibility using numerical
simulation on simple cubic lattices of size $L^3$, with $32 \leq L \leq 480$.
The programs and algorithms used are essentially the same as the ones
described in the previous section. The basic procedure is to generate
${\cal M}$ independent runs for each lattice size and compute the fraction
of those which percolate after the bootstrap condition is used and a
stable configuration is reached. The values of ${\cal M}$ are the same as
those used in the calculation of $y_h$ (see previous section).

\begin{figure}
\epsfxsize=9.5cm
\begin{center}
\leavevmode
\epsffile{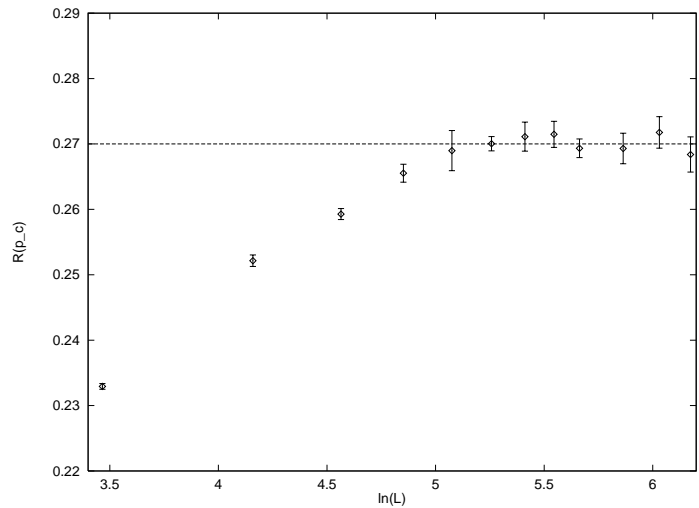}
\caption{Graph of $R(p_c) \times \log(L)$, for $32 \leq L \leq 480$. The
linear size $L$ is always a multiple of 32. The dotted
line is a guide to the eye and indicates the value 0.270, taken as $R(p_c)$
in the limit $L \rightarrow \infty$.}
\label{rpc} 
\end{center}
\end{figure}

	The results are depicted in Figure \ref{rpc}:
we can infer that $R(p_c)=0.270 \pm 0.005$, where the error bar is a rough
estimate. There are two previous calculation of $R(p_c)$ for the uncorrelated
site percolation on the cubic lattice with free boundary conditions: they
lead to $R(p_c)=0.265 \pm 0.005$ \cite{Lin} and $R(p_c)=0.28$ \cite{Acha}.
Our value agrees, within the numerical error, with the first one but the
result of Reference \cite{Acha} cannot be ruled out. It is then
reasonable to infer that usual percolation and $m=3$ bootstrap percolation
on the cubic lattice belong to the same universality class, also in
what regards the critical spanning probability.

\section{Summary} \label{summ}

	We calculate, using numerical simulations and finite-size
scaling techniques, the critical parameters of the bootstrap percolation
model with $m=3$ on the cubic lattice, using finite lattices of size $L^3$, 
with
$32 \leq L \leq 480$. Our evaluations of $\nu$ and $\beta$ strongly support
the conclusion that usual percolation and the model we study belong to the
same universality class. This result disagrees with previous calculations
\cite{kogut,as};
we believe this is due to the small sizes used in those works. To support
this assumption, we note that the finite-size scaling assumptions hold for
lattices of linear size $L \geq 128$, which are higher then the sizes
studied in previous works.

	The critical spanning probability, $R(p_c)$, is also calculated. 
It has been shown that this quantity shows some degree of universality, but, 
to the best of our knowledge, no study concerning correlated percolation
models have been done so far. Our result for bootstrap
percolation with $m=3$ on the cubic lattice, $R(p_c)=0.270 \pm 0.005$, 
is, within the numerical accuracy,
the same value as for usual percolation with free boundary conditions.
Therefore, we can infer that $R(p_c)$ is not sensitive to short range
correlations and even to some long range correlations, like the one studied
in this paper.

\vskip 0.7cm

	We would like to thank Dr. D. Stauffer for fruitful 
discussions at early stages of this work and for a critical reading of the 
manuscript.

\end{document}